\documentclass[dvips]{article}

\usepackage{icrc2011}

\title{A statistical model to explain the gamma-ray variability and flares
of the Crab nebula}

\newcommand{\etal}{\MakeLowercase{\textit{et al. }}} 
\shorttitle{Yuan \etal Variability and flare of Crab nebula}

\authors{Qiang Yuan$^{1}$, Peng-Fei Yin$^{1}$, Xue-Feng Wu$^{2,3}$,
Xiao-Jun Bi$^{1}$, Siming Liu$^{4}$, Bing Zhang$^{5}$}
\afiliations{$^1$Key Laboratory of Particle Astrophysics, Institute of 
High Energy Physics, Chinese Academy of Sciences, Beijing 100049, China\\ 
$^2$Purple Mountain Observatory, Chinese Academy of Sciences, Nanjing
210008, China\\
$^3$Joint Center for Particle Nuclear Physics \& Cosmology (J-CPNPC),
Nanjing 210093, China\\
$^4$Key Laboratory of Dark Matter and Space Astronomy, Purple Mountain 
Observatory, Chinese Academy of Sciences, Nanjing 210008, China\\
$^5$Department of Physics and Astronomy, University of Nevada Las Vegas,
Las Vegas, NV 89154, USA}
\email{yuanq@mail.ihep.ac.cn}

\abstract{Recently the AGILE and Fermi/LAT detectors uncovered giant 
$\gamma$-ray flares from the Crab nebula. The duration of these flares
is a few days. The Fermi/LAT data with monthly time binning further 
showed significant variability of the synchrotron tail of the emission, 
while the inverse Compton component was stable. The simultaneous or 
follow-up observations in X-ray, optical, infrared and radio bands did 
not find significant flux variation. Based on these observations, we 
propose that the $\gamma$-ray variability and flares are due to 
statistical fluctuations of knots that can accelerate electrons to 
$\sim$PeV energies. The maximum achievable energy of electrons is adopted 
to be proportional to the size of the knot, which is assumed to follow 
a power-law distribution. Thus the low energy electron flux will be 
stable due to the large number of small knots, while the high energy 
electron flux may experience large fluctuations. Monte Carlo realization 
of such a picture can reproduce the observational data quite well given 
proper model parameters.
}

\begin{document}
\maketitle

\section{Introduction}

The Crab nebula is one of the most famous source in the sky, and is
the most typical multi-band laboratory. It has been widely studied in 
all wavelengths, from radio to very high energy (VHE) $\gamma$-rays. 
The broadband non-thermal emission can be well modeled with a leptonic
scenario, with the synchrotron radiation from the radio to GeV
$\gamma$-rays and inverse Compton (IC) radiation above $\sim$GeV
\cite{Atoyan1996}. The transition from synchrotron to IC component was
clearly seen by CGRO/EGRET \cite{Kuiper2001} and Fermi/LAT \cite{Abdo2010}.
Two populations of electrons, i.e., the radio electrons and wind electrons
are required to fit the data \cite{Atoyan1996}. The overall emission from 
the nebula seems to be stable in radio, optical, X-ray and VHE $\gamma$-ray 
bands, thus it was thought to be a ``standard candle'' and was often used 
to calibrate the detectors.

However, years ago people began to find that the MeV-GeV emission of Crab
nebula was actually variable instead of stable, from the COMPTEL and
EGRET observations \cite{Much1995,deJager1996}. Furthermore the detailed 
images of the Crab nebula in optical and X-ray bands also indicated 
dynamical structures at small scales. The HST observation revealed highly
variable wisps and knots in the nebula \cite{Hester1995}. X-ray 
observations by ROSAT and Chandra uncovered a jet-torus structure 
of the inner nebula, which was also dynamically variable 
\cite{Hester1995,Weisskopf2000}. Recently in September 2010, the AGILE 
X-ray satellite detected a $\gamma$-ray flare above $100$ MeV from 
Crab nebula, with a flux enhancement of $\sim3$ times higher than the 
average one and a duration of about 3 days \cite{Tavani2011}. This flare 
was confirmed by the Fermi/LAT detector, with an even higher flux 
enhancement \cite{Abdo2011}. The search for the archive data of AGILE
and Fermi/LAT further unveiled other flares, in October 2007 for AGILE 
and in February 2009 for Fermi/LAT \cite{Tavani2011,Abdo2011}, which
may indicate that such flare events occur with a timescale about one
year. Most recently in April 2011, AGILE detector observed another 
super-flare from Crab nebula \cite{Striani2011}. Furthermore the 
monthly binned light curve of Fermi/LAT data showed that the synchrotron 
component is variable, however, no variation was found for the high 
energy IC component \cite{Abdo2011}.

There were many other simultaneous or follow-up observations for the 
flare in September 2010 in other wavelength bands. In the VHE $\gamma$-ray 
energies, ARGO-YBJ collaboration reported a detection of a flux enhancement
around TeV with a possibly longer duration (ATel \#2921). The results
from MAGIC and VERITAS telescopes with limited exposure time, however,
did not find any flux variation during the flare phase (ATel \#2967, 
\#2968). In X-ray, optical, infrared and radio bands, no significant 
flux enhancement was discovered either (ATel \#2856, \#2858, \#2866,
\#2867, \#2868, \#2872, \#2882, \#2889, \#2893, \#2903). 

Some theoretical models are proposed to explain the flare event 
\cite{Komissarov2011,Bednarek2011}. In \cite{Yuan2011} we proposed
that the $\gamma$-ray variability and flare were due to the statistical
fluctuation of the acceleration units which were responsible for the 
highest energy electrons. It is natural to expect that events that 
can generate the highest energy electrons are rarer, and therefore 
would suffer from the largest fluctuations. The lower energy electron
spectrum should be steady-like because many more accelerators can 
contribute to them simultaneously. Since the synchrotron $\gamma$-rays 
are produced by the highest energy electrons ($\sim$PeV), while the 
synchrotron emission at lower energies and IC emission at higher energies
are produced by lower energy electrons, we can easily explain why we only
detect variability in MeV-GeV $\gamma$-rays, neither in lower nor in 
higher bands.

\section{Model}

We assume that electrons are accelerated in a series of knots, which
may relate with the magnetic hydro-dynamic turbulence of the plasma. 
For each knot the output energy spectrum of electrons is a power-law 
function $F_i(E) \propto E^{-\alpha_e}$ with a maximum electron energy
$E_{\rm max}^i$, which is proportional to the size of the $i$th knot.
It is further assumed that the size of knots has a power-law distribution 
$P(r_i)\propto r_i^{-\beta}$. The total electron spectrum can be got
through adding the contribution from all the knots together.

The comoving system synchrotron spectrum from a knot with size $r_i$ is 
$\nu'F_{\nu'}\propto\nu'^{-\alpha_{\nu}} \exp(-\nu'/\nu'_{\rm max})$, 
where $\nu'_{\rm max} \propto(E_{\rm max}^i)^2 \propto r_i^2$, and 
$\alpha_{\nu}=(\alpha_e-3)/2$ is the synchrotron spectral index, with
$\alpha_e$ being the power-law index of the electron distribution.
Changing to the observer's frame, there will be a frequency shift
$\nu'\rightarrow\delta\nu'$ and a flux enhancement $F_{\nu'}\rightarrow
\delta^3F_{\nu'}$, where $\delta$ is the Doppler factor of the knot.

It was shown that the maximal energy of synchrotron emission in the 
magnetic field dominated acceleration regime is $\sim 70$ MeV, which
is due to the fact that the electron may lose most of its energy in
one cycle of the Larmor motion \cite{Blumenthal1970}. For smaller 
accelerators, the maximum energy is further limited by the size of 
the accelerator. The observational flare of Crab nebula has energies
higher than GeV, which means that either there is Doppler boost of
the flare event, or the acceleration of the electrons is not the
shock-like scenario \cite{Abdo2011}. In order to explain the high 
energy photons of the flare event, we therefore employ a mild Doppler 
boost factor. Only knots with large enough size and large enough Doppler 
factor can contribute to high energy synchrotron radiation to explain 
the flares observed by Fermi/LAT.

The cooling of the electrons need to be considered. The cooling time
of the electrons is energy dependent. For synchrotron photon with energy
$\epsilon$ the cooling time is $t_c\simeq 1.5(B/{\rm mG})^{-1.5}
(\epsilon/{\rm keV})^{-0.5}\delta^{-0.5}$ yr. If we adopt an average
magnetic field of the Crab nebula $B\approx0.1$ mG, a mildly Doppler
factor $\delta\sim$1, the electrons correspond to $\epsilon\sim2.5$
eV photons will have a cooling time comparable to the age of the nubela
$t_{\rm age}\approx10^3$ yr. If the production of the knots is continous,
the equilibrium electron spectral index should be\footnote{Note that here
the equilibrium spectrum actually represents a series of knots with the
same sizes, instead of a single one.} $\alpha_e=\alpha_e^
{\rm inj}+1$, for electrons with cooling time shorter than $t_{\rm age}$.
For the electrons with lower energies, whose cooling time is even longer
than the age of the nebula, there will be no cooling at all and the
injection spectrum will keep unchanged. Here we will focus on the high
energy part, e.g., the synchrotron spectra from X-ray to $\gamma$-ray
band, we will take a cooled spectrum of electrons as input of the model.

\section{Monte Carlo simulation}

\begin{figure}[!htb]
\centering
\includegraphics[width=\columnwidth]{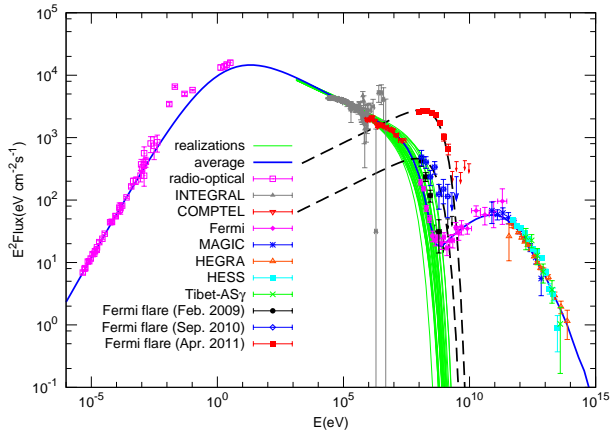}
\caption{Spectral energy distribution of the Crab nebula. Thin-green-solid 
lines: simulated synchrotron spectra in $30$ realizations; thick-blue-solid: 
fit of the average multi-wavelength emission; black-dashed: examples of
unusual events which may be responsible for the flares. Data are from:
radio-optical \cite{Macias-Perez2010}, INTEGRAL \cite{Jourdain2009},
COMPTEL \cite{Kuiper2001}, Fermi/LAT \cite{Abdo2010}, 
MAGIC \cite{Albert2008}, HEGRA \cite{Aharonian2004}, 
HESS \cite{Aharonian2006}, Tibet-AS$\gamma$ \cite{Amenomori2009},
and Fermi/LAT flares \cite{Abdo2011,Buehler2011}.}
\label{fig:f1}
\end{figure}

We realize the above picture by a Monte-Carlo simulation. The injection 
spectrum of electrons from the knots is adopted as $\alpha_e^{\rm inj}=
1.6$, which corresponds to the fit to the radio-optical spectrum of the
nebula \cite{Meyer2010}. After taking into account the cooling effect 
the electron spectrum is $\alpha_e=2.6$. We directly generate the 
synchrotron spectrum instead of the electron spectrum. The corresponding 
synchrotron spectral index is then $\alpha_{\nu}=-0.2$. The maximum
energy of the synchrotron photon in the knot comoving system is 
proportional to $r_i^2$. We normalize $\nu'_{\rm max}$ of the largest
knot(s) to be 55 MeV to account for the constraints of cooling effect
during the acceleration. The normalization of the synchrotron emmisivity 
from each knot is adopted a volume-proportional factor $r_i^3$.
As for the Lorentz factors of the knots, we assume a Gaussian distribution 
with the mean value $\Gamma=2.0$ and a standard deviation $\sigma=0.25$. 
Such a Lorentz factor is consistent with the upper limit of the typical 
velocity of the jet \cite{Abdo2011}. The angle $\theta$ between the knot 
motion and the line-of-sight is assumed to be randomly distributed. 
For $3\sigma$ range of $\Gamma$ we find the Doppler factor 
$\delta=1/\Gamma(1-v\cos\theta)$ is in the range $[0.18,5.5]$.
Finally the power-law index of knot size distribution is adopted as 
$\beta\approx4.8$. After adding the contribution from all the knots 
with different sizes together we get the total synchrotron spectrum
$\nu'F_{\nu'}\propto\nu'^{-0.2}$, which is can well reproduce the
observed spectra of Crab nebula from optical to MeV $\gamma$-ray band,
as shown in Figure \ref{fig:f1}.

The simulated spectra together with the observational data are shown
in Figure \ref{fig:f1}. For comparison we show the fitting results
to the broad band data with the thick blue line, with a broken
power-law electron spectrum. The electron spectral indices are 
$\beta_1/\beta_2=1.60/3.45$ below/above the break Lorentz factor 
$\gamma_{\rm br}=1.1\times 10^6$. The high energy cutoff is adopted 
as a super-exponential behavior $\gamma^{-\beta_2}\exp[-(\gamma/
\gamma_{\rm cut})^{\delta}]$, with $\gamma_{\rm cut}=5\times 10^9$ 
and $\delta=2.0$. The magnetic field $B=124$ $\mu$G, which is a 
constant in the all nebula. This result is similar to the model 
invoking two population electrons as introduced in \cite{Atoyan1996} 
and \cite{Meyer2010}. The average of the simulated spectra is similar
to the steady state fit, and is consistent with the long term Fermi/LAT 
and COMPTEL data. However, there is scattering of the simulated spectra,
which reflects the observational variability of the $\gamma$-ray
spectrum. In the X-ray band, the fluctuation is very small, which is
close to a steady state emission.

In this model, the IC component should not vary significantly. This is 
because the IC photons around TeV energies are primarily produced by the 
low energy electrons. For instance for $\sim100$ TeV electrons the 
synchrotron photon energy is typically $0.016(B/{\rm mG})(E/{\rm TeV})^2$ 
keV $\sim16$ keV for $B\sim 0.1$ mG, which suffers from very small 
fluctuation as shown in Figure \ref{fig:f1}. The corresponding IC 
photon energy is $\epsilon_{\rm IC}\sim\max(\gamma^2\epsilon,
\gamma m_e)$. For background photon energy $\epsilon\sim$eV the IC 
photon energy is as high as 100 TeV. Even for the cosmic microwave 
background photon $\epsilon\sim 10^{-3}$ eV we still have 
$\epsilon_{\rm IC}\sim 10$ TeV. Therefore the fluctuation of the IC 
component should be indeed small. This result is consistent with the 
Fermi/LAT observations \cite{Abdo2011}.

The two large flares in September 2010 and April 2011 are not well 
reproduced in the simulation. We show two illustrations of these
events with proper parameters in Figure \ref{fig:f1}. For the September 
2010 flare the model parameters are $\delta=5.5$, $\nu'_{\rm max}=70$ MeV.
These parameters are reasonable in the present frame. However, for the
April 2011 flare, we may need $\delta=8.0$, $\nu'_{\rm max}=70$ MeV,
which seem to be very extreme. Such a flare can be regarded as an event 
with very small probability. 

\begin{figure}[!htb]
\centering
\includegraphics[width=\columnwidth]{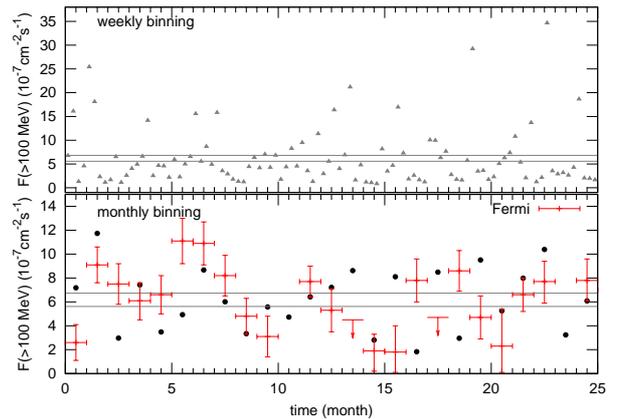}
\caption{Light curves of the simulated synchrotron flux above $100$
MeV. The upper panel shows the results with weekly bin and the
bottom panel for monthly bin. The average flux of the realizations
is normalized to the observational one $\sim 6.2\times 10^{-7}$ 
cm$^{-2}$ s$^{-1}$ \cite{Abdo2011}, as shown with the horizontal
lines. Also plotted in the lower panel are the Fermi/LAT observational 
data with monthly bin.}
\label{fig:f2}
\end{figure}

To investigate the fluctuations of emission in detail, we calculate the 
light curves of the simulated photon fluxes. Since there is purely 
statistical fluctuation of the fluxes, we can adopt
independent realizations to represent each time bin. The length of
the time bin is proportional to the number of knots. We need a 
normalization of the absolute time scale, which is adopted that in 
one year there is an enhancement in flux by at least a factor of 5-6 
for weekly bins, which may be responsible for a flare. We plot the weekly 
and monthly bin lightcurves in the upper and lower panels of Figure 
\ref{fig:f2}, respectively. The Fermi/LAT observed monthly light curve 
\cite{Abdo2011} is also plotted in the lower panel of Figure \ref{fig:f2}
for comparison. It can be seen that the predicted monthly variability 
scale of the simulated results is very similar to that observed by 
Fermi/LAT. 

\begin{figure*}[!htb]
\centering
\includegraphics[width=\columnwidth]{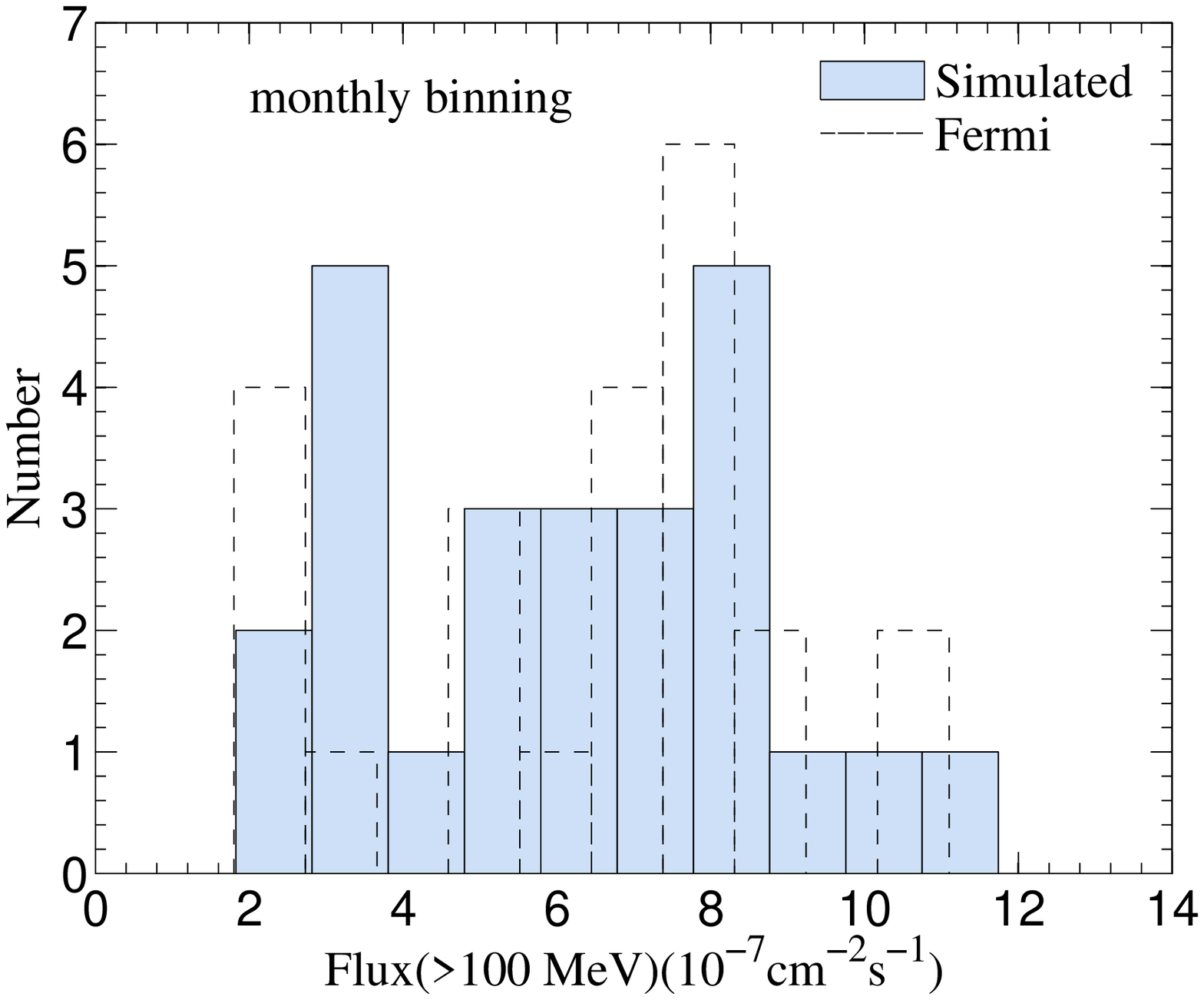}
\includegraphics[width=\columnwidth]{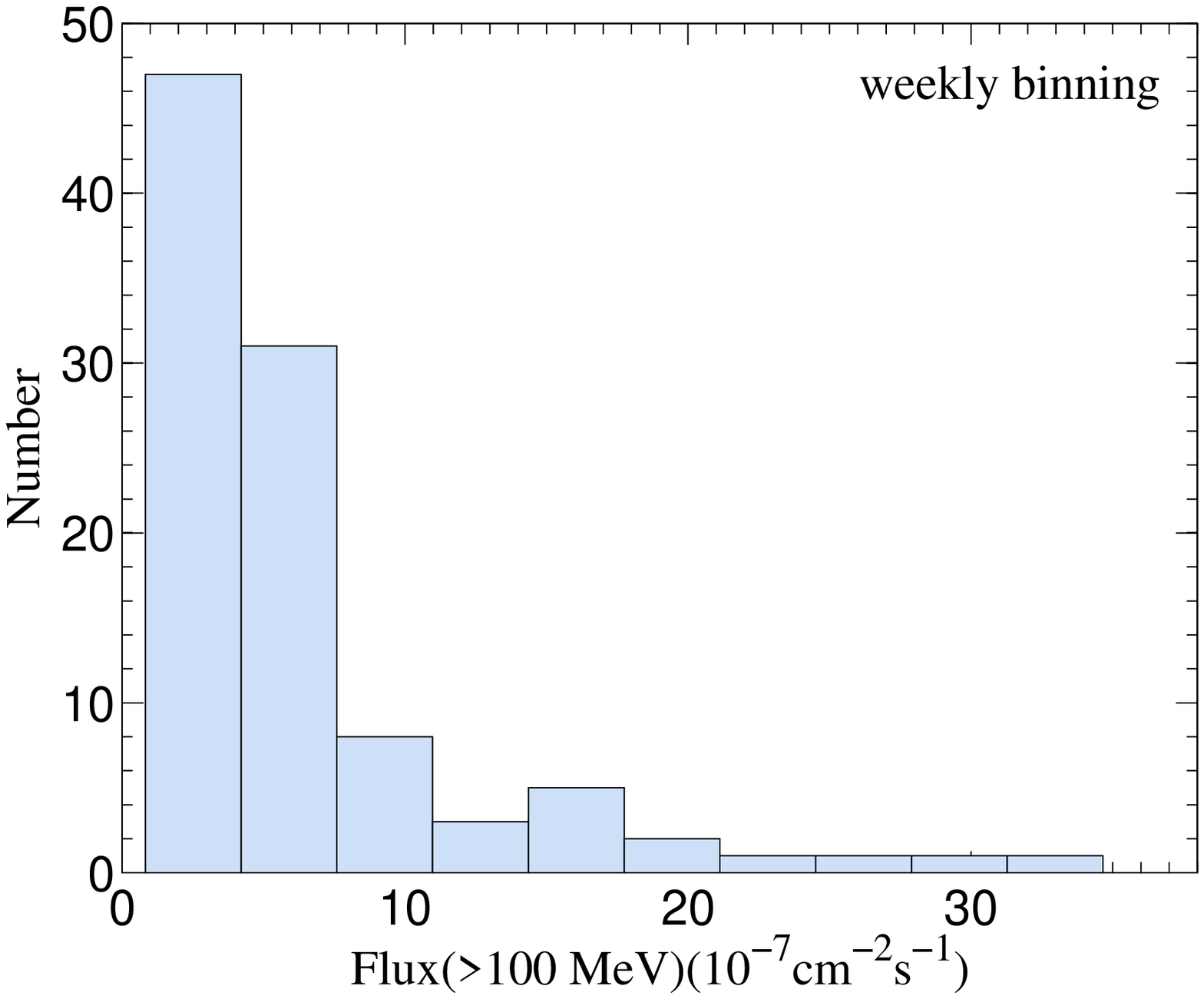}
\caption{Histograms of the synchrotron flux distribution
above 100 MeV, for monthly bin (left) and weekly bin
(right), respectively. In the left panel the Fermi/LAT
observation result is also shown (dashed histogram).}
\label{fig:f3}
\end{figure*}

Figure \ref{fig:f3} shows the histogram of the flux distribution.
For the monthly bin results, we can see clearly the similarity 
between the simulation and the observational data. In the right panel 
of Figure \ref{fig:f3} the histogram of flux distribution for weekly 
bin is shown. We can see that there is a significant concentration 
toward low-flux, which is distinct from a symmetric distribution 
around the average value. The detailed analysis of Fermi/LAT 
data can test such a prediction.

\section{Conclusion}

In this work we propose that the fluctuations of the electron spectra 
at the highest energy end ($\sim$PeV) are responsible for the variability 
of the synchrotron tail in GeV $\gamma$-ray band. The electrons are 
thought to be accelerated in a series of knots, with a size distribution 
$P(r_i)\propto r_i^{-\beta}$ and a distribution of the Doppler factor. 
The maximal achievable energy of the electrons in the co-moving frame 
is assumed to be proportional to the size of the knots. Thus the very 
high energy electrons are generated by the very rare knots with both
large sizes and high Doppler boosts, and hence, suffer from large 
fluctuations. On the other hand, the low energy electrons can be 
accelerated by many smaller knots, and the fluctuations are smoothed
out. Therefore only the variability in MeV-GeV $\gamma$-ray band is
observed, because it is the direct reflection of the highest energy
electrons. The lower energy synchrotron component and the higher
energy IC component are relatively stable due to the less fluctuation
of the low energy electrons. The expected variability of the monthly 
bin fluxes above 100 MeV are well consistent with that observed by
Fermi/LAT. The two large $\gamma$-ray flares can also be naturally
accounted for without additional assumptions.

It has been revealed recently that in the X-ray band the Crab nebula
is actually no longer ``standard candle'', but experiences quasi-periodic
variability with a level of several percentage \cite{Wilson-Hodge2011}.
Such large scale variability may be due to the overall energy injection
of the nebula or the inhomogeneous plasma flow induced magnetosonic 
waves \cite{Wilson-Hodge2011}. The model proposed here may not be
responsible for such variability. However, it would be interesting
to investigate the statistical fluctuations of the X-ray fluxes
in different energy ranges after removing the large scale variability,
and compare with our model prediction.

One thing we should keep in mind is that in the present model we
do not expect the variability of the IC component. This conclusion
is consistent with the Fermi/LAT data, however, in a relatively long
time scale (monthly bin). During the flare phase, we note that the
ARGO-YBJ experiment detected a potential flux enhancement in TeV
energy range (ATel \#2921). Although it was not confirmed by the
Cherenkov telescopes MAGIC (ATel \#2967) and VERITAS (ATel \#2968), 
we should note that the exposure time of MAGIC (1 hr) and VERITAS 
(2 hr) is very limited. It is thus very important to further monitor
the Crab nebula in VHE energy band by e.g., Tibet-MD \cite{Amenomori2007}
or LHAASO \cite{Cao2010}, to search for the variability, which can be 
a crucial test of the current model.

This work is supported by the Natural Sciences Foundation of China
grants 10773011, 11075169, 10633040 and 10921063, the 973 project 
grants 2010CB833000 and 2009CB824800, the NSF grant AST-0908362 and 
NASA grants NNX10AD48G and NNX10AP53G at UNLV.

\clearpage


\begin{thebibliography}{}

\bibitem{Atoyan1996}
{Atoyan}, A.~M., {Aharonian}, F.~A., 1996, MNRAS, {\bf 278}, 525

\bibitem{Kuiper2001}
{Kuiper}, ~L. et al., 2001, A\&A, {\bf 378}, 918

\bibitem{Abdo2010}
{Abdo}, ~A. et al., 2010, ApJ, {\bf 708}, 1254

\bibitem{Much1995}
{Much}, ~R. et al., 1995, A\&A, {\bf 299}, 435

\bibitem{deJager1996}
{de Jager}, ~O.~C. et al., 1996, ApJ, {\bf 457}, 253

\bibitem{Hester1995}
{Hester}, ~J.~J. et al., 1995, ApJ, {\bf 448}, 240

\bibitem{Weisskopf2000}
{Weisskopf}, ~M.~C. et al., 2000, ApJ, {\bf 536}, L81

\bibitem{Tavani2011}
{Tavani}, ~M. et al., 2011, Science, {\bf 331}, 736

\bibitem{Abdo2011}
{Abdo}, ~A. et al., 2011, Science, {\bf 331}, 739

\bibitem{Striani2011}
{Striani}, ~E. et al., ArXiv e-prints:1105.5028

\bibitem{Komissarov2011}
{Komissarov}, ~S.~S., {Lyutikov}, ~M., 2011, MNRAS, {\bf 414}, 2017

\bibitem{Bednarek2011}
{Bednarek}, ~W., {Idec}, ~W., 2011, MNRAS, {\bf 414}, 2229

\bibitem{Yuan2011}
{Yuan}, ~Q. et al., 2011, ApJ, {\bf 730}, L15

\bibitem{Blumenthal1970}
{Blumenthal}, ~G.~R., {Gould}, ~R.~J., 1970, Reviews of Modern Physics, 
{\bf 42}, 237

\bibitem{Meyer2010}
{Meyer}, ~M., {Horns}, ~D., {Zechlin}, ~H., 2010, A\&A, {\bf 523}, A2

\bibitem{Macias-Perez2010}
{Macias-Perez}, ~J.~F. et al., 2010, ApJ, {\bf 711}, 417

\bibitem{Jourdain2009}
{Jourdain}, ~E., {Roques}, ~J.~P., 2009, ApJ, {\bf 704}, 17

\bibitem{Albert2008}
{Albert}, ~J. et al., 2008, ApJ, {\bf 674}, 1073

\bibitem{Aharonian2004}
{Aharonian}, ~F. et al., 2004, ApJ, {\bf 614}, 897

\bibitem{Aharonian2006}
{Aharonian}, ~F. et al., 2006, A\&A, {\bf 457}, 899

\bibitem{Amenomori2009}
{Amenomori}, ~M. et al., 2009, ApJ, {\bf 692}, 61

\bibitem{Buehler2011}
{Buehler}, ~R., Fermi Symposium, Roma 2011

\bibitem{Amenomori2007}
{Amenomori}, ~M. et al., ArXiv e-prints:0710.2757

\bibitem{Cao2010}
{Cao}, ~Z., 2010, Chinese Physics C, {\bf 34}, 249

\bibitem{Wilson-Hodge2011}
{Wilson-Hodge}, ~C.~A. et al., 2011, ApJ, {\bf 727}, L40

\end{thebibliography}
\end{document}